\documentclass[aps,prl,twocolumn,groupaddress,superscriptaddress,nofootinbib,showkeys,showpacs,altaffilletter]{revtex4-1}

\usepackage{graphicx}
\usepackage{dcolumn}
\usepackage{amssymb}
\usepackage{amsmath}
\usepackage{amsfonts}
\usepackage{amsbsy}
\usepackage{color}
\usepackage{rotating}
\usepackage[english]{babel}
\usepackage{soul}

\begin{document}

\title{Measuring the speed of light with Baryon Acoustic Oscillations}

\date{\today}

\author{Vincenzo Salzano}
\affiliation{Institute of Physics, University of Szczecin, Wielkopolska 15, 70-451 Szczecin, Poland}
\author{Mariusz P. D\c{a}browski}
\affiliation{Institute of Physics, University of Szczecin, Wielkopolska 15, 70-451 Szczecin, Poland}
\affiliation{Copernicus Center for Interdisciplinary Studies, S{\l}awkowska 17, 31-016 Krak{\'o}w, Poland}
\author{Ruth Lazkoz}
\affiliation{Fisika Teorikoaren eta Zientziaren Historia Saila, Zientzia eta
Teknologia Fakultatea, \\ Euskal Herriko Unibertsitatea, 644 Posta Kutxatila,
48080 Bilbao, Spain}


\begin{abstract}
In this letter we describe a new method to use Baryon Acoustic Oscillations (BAO) to derive a constraint on the possible variation of the speed of light. The method relies on the fact that there is a simple relation between the angular diameter distance $(D_{A})$ maximum and the Hubble function $(H)$ evaluated at the same maximum-condition redshift, which includes speed of light $c$. We note the close analogy of the BAO probe with a laboratory experiment: here we have $D_{A}$ which plays the role of a standard (cosmological) ruler, and $H^{-1}$, with the dimension of time, as a (cosmological) clock. We evaluate if current or future missions such as \textit{Euclid} can be sensitive enough to detect any variation of $c$.
\end{abstract}




\maketitle

\textit{Introduction.} The present standard cosmological model (General Relativity plus Standard Model of Particle Physics) is very successful in explaining many observational facts of our Universe, but not everything \citep{Perivolaropoulos}. 
Beyond observational inconsistencies, deeper and more profound problems are also in play (singularity, horizon and curvature problems; dark energy; multiverse question) which motivate the search for alternative solutions. One very interesting branch of such investigations is the assumption that the physical constants of our known physics are not really constant, but might have been varying in the past (and possibly still are varying). Seminal ideas for such scenario are quite old \citep{VSL_old}, but only recently (for a review, see \citep{Uzan2011}) they have come to interest for their intriguing relation with a more profound level of knowledge of the physics of the universe.

Theories of varying speed of light (VSL) have attracted much interest in the last years: while there is still much controversy about them \citep{VSL_controversy}, many theoretical works have been produced \citep{VSL_theory,Magueijo2003}, but a reliable study of their application to observational data is still missing \citep{VSL_observations}. Here, we will focus on the possibility to extract constraints on a possible VSL theory by using BAO \citep{BAO_review}. We want to stress that our approach will be \textit{completely cosmological-model-independent}: no background cosmology will be necessary and we will - only - need observational data and related observational errors as essentials.

\begin{figure}[htbp]
\includegraphics[width=8.3cm]{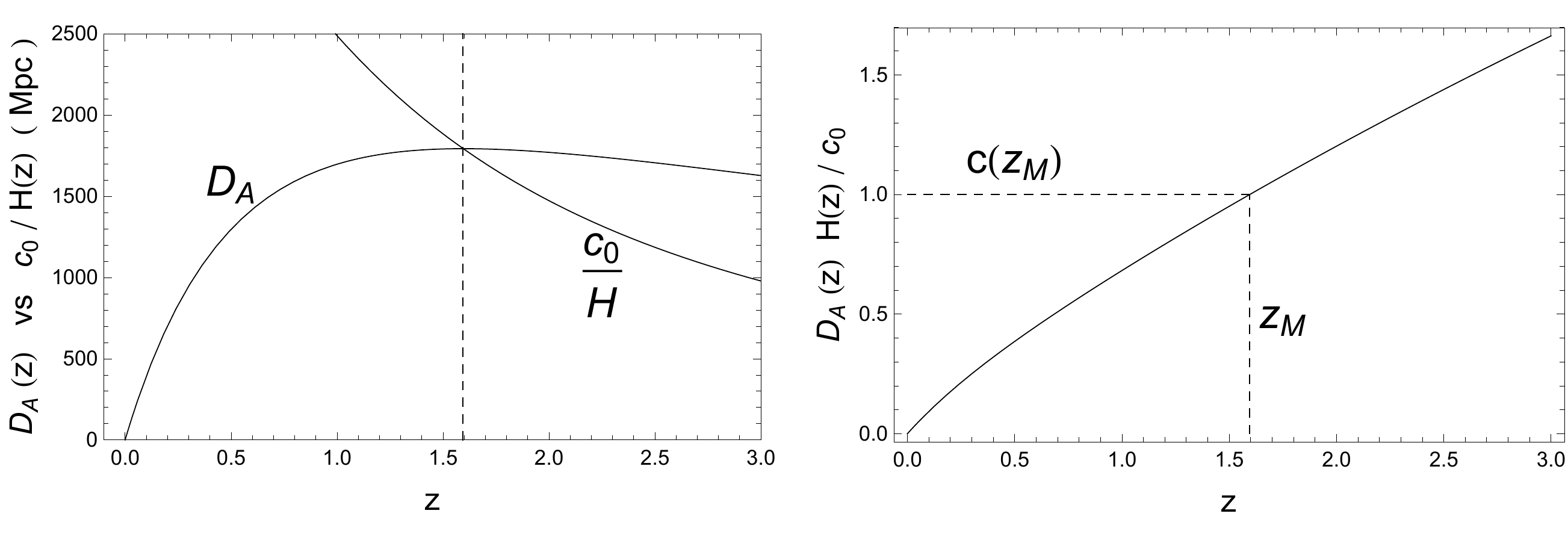}
\caption{Method implementation: maximum redshift detection \textit{left panel}; speed of light measurement \textit{right panel}.}\label{fig:DAH}
\end{figure}

\textit{Theoretical basis.} The possibility to constrain VSL theories from BAO resides on the definition of one of the quantities that are generally measured with BAO, the angular diameter distance, $D_{A}(z)$, where $z$ is the redshift. It is well known \citep{Weinberg} that $D_{A}(z)$ rises up to a maximum at some redshift $z_{M}$, and then starts to decline. The exact value for $z_{M}$ depends on the cosmological model; using the $w+w_{a}$ \citep{CPL} \textit{Planck+WMAP+highL+BAO} bestfit \citep{Planck_Archive} and varying its parameters in their $1\sigma$ confidence intervals \citep{footnote1},
we have checked that $z_{M}$ lies in the range $[1.4,1.8]$ for more than $95\%$ of $10^{4}$ random cosmological models. 
This range is consistent with a flat dust Friedmann universe (no cosmological constant), $z_{m} = 1.25$, and with cosmic strings, $z_{m} = e - 1$, ($e$ the Euler number) \citep{MPD87}. Given the large degeneracy between the cosmological parameters, $z_{M}$ is of no real use to constrain dark energy. But $D_{A}$ and $z_{M}$ are very interestingly related: the condition for the maximum, $\partial D_{A}(z) / \partial z = 0$, evaluated at $z_{M}$, implies $D_{A}(z_{M}) = c_{0}/H(z_{M})$, where $H(z)$ is the Hubble function (left panel in Fig.~\ref{fig:DAH}). From now on, we will define $c_{0} \equiv 299792.458$ km s$^{-1}$ as the value of the speed of light, which is assumed constant by the standard scenario, or equal to the speed of light evaluated \textit{here and now} in a VSL theory.

This relation is very important for determining the value of $z_{M}$. Using only $D_{A}$ for this purpose would be problematic. The combination of a number of effects smear out the profile of $D_{A}(z)$: the quite large plateau at about $z_{M}$; measuring $D_{A}(z)$ from just a few redshift bins from a BAO survey; the errors plus the intrinsic dispersion of the measurements. The final consequence is the practical impossibility to determine the location of the maximum. But, from BAO surveys (in particular from future surveys), it will be possible to extract independently both the tangential mode, $D_{A}(z)$, and the radial mode, $c_{0}/H(z)$. Comparing observational data from both $D_{A}(z)$ and $c_{0}/H(z)$ it will be possible, in principle, to constraint the value of $z_{M}$ better, because instead of searching for the maximum in $D_{A}(z)$, one can search for the redshift at which $D_{A}(z_{M}) = c_{0}/H(z_{M})$ holds.

Even so, anyway, one should expect better but still far-from-precise constraints on $z_{M}$ for the same reasons listed above (binned data, errors, \dots). But one can employ some cosmological-model-independent method to extract information from data. Examination of the rich literature about this \citep{Reconstruction_Methods} suggests that Gaussian Processes (GPs) \citep{GP,Seikel2012} are very well suited to our needs, and we have employed them to reconstruct $D_{A}(z)$ and $c_{0}/H(z)$ in order to find $z_{M}$.

Thus, the application of GPs to BAO modes yields $D_{A}(z)$ and $c_{0}/H(z)$ numerically-reconstructed as smooth analytical functions, ready for evaluation at whatever redshift value one may need. GPs are also very helpful because they incorporate in a very natural and straightforward way correlations between data (non-diagonal covariance matrix). In fact that is the kind of situation we are concerned with, because the tangential and radial BAO modes are known to be correlated \citep{SeoEisenstein2007}.  The sets of GP-reconstructed BAO modes can eventually be employed in a numerical algorithm to estimate $z_{M}$ and its error.

The determination of $z_{M}$ is vital for our main purpose here, namely, the detection of variability of $c$. Eventually, the quantity $D_{A}(z) \cdot H(z)$ can easily be obtained from the original data by applying GPs. Then, upon evaluation at $z_{M}$ it will be straightforward to check whether $D_{A}(z_{M}) \cdot H(z_{M}) = c_{0}$ (or $D_{A}(z_{M}) \cdot H(z_{M})/c_{0} = 1$, if we want to express it in terms of primarily observed quantities). Thus $D_{A}(z) \cdot H(z)$, evaluated at the maximum-point redshift $z_{M}$, will unequivocally give the value of the speed of light at this redshift (right panel in Fig.~\ref{fig:DAH}). We want to stress here the main strong point of this relation: changing the cosmological model and the VSL theory will both change the value of $z_{M}$ and the general profile of $D_{A}(z)$ and $H(z)$. One could argue that a VSL theory might be constrained directly from the observations, with no need of any alternative method. But in that case we would have uncertainties adding up, from the cosmological and the VSL theory, which we both ignore. Moreover, they are degenerate, thus leaving impossible to determine with good accuracy any of them. Our method is different, because \textit{the relation between $z_{M}$ and $D_{A}(z_{M}) \cdot H(z_{M})$ will be always the same, namely, given any cosmological model, we will always have $D_{A}(z_{M}) \cdot H(z_{M}) = c_{0}$.}

In a standard context where the speed of light is not expected to change, combining the errors on $z_{M}$ with the errors on $D_{A}(z) \cdot H(z)$ will measure $c_{0}$ with some error. Actually the measurement of $c_{0}$ is assumed to be exact and is used as the reference rule for the definition of the meter \citep{velocity_c_1}; the best measurement having a relative error $\sim 10^{-9}$ \citep{velocity_c_2}. This precision, obtained with laser interferometry, is largely out of the possibility of a cosmological measurement. But if we assume a VSL, i.e. the existence of an - up to now unknown - function $c(z)$ (with the limit $c(z\rightarrow0) \equiv c_{0}$), then BAO modes would be revealing to us that $D_{A}(z_{M}) \cdot H(z_{M}) = c(z_{M})$, where $c(z_{M}) \neq c_{0}$ is the value of the speed of light at redshift $z_{M}$. Deviations from $c_{0}$, $\Delta c \equiv c(z_{M})- c_{0}$, if any, can be of whatever order possible, not necessarily as small as $10^{-9}$.

Let us summarize our method: we take $(D_{A}(z), H(z))$; then use them to calculate the maximum-point redshift $z_{M}$; and finally evaluate $D_{A}(z) \cdot H(z)$ at $z_{M}$ so as to estimate $\Delta c$ and detect possible VSL.

\textit{Implementation.} In \citep{FontRibera2014} the constraints on $D_{A}$ and $H$ from many on-going and future BAO surveys are analyzed. Among them, the best results are from the ESA mission \textit{Euclid} \citep{EUCLID_1,EUCLID_2}. In Table~6 of \citep{FontRibera2014} the percentage errors on $D_{A}/s$ and $H \cdot s$ ($s$ is the sound horizon at decoupling \citep{FontRibera2014}) for $15$ redshift bins in the redshift range $[0.6;2.1]$  covered by \textit{Euclid}  are given (the bin width is $0.1$). In order to produce some mock data to work with, we need a fiducial cosmological model; we choose the baseline model ($\Lambda$CDM) from \textit{Planck+WMAP+highL+BAO} (last column of Table~5 in \citep{Planck_26}), the same used by \citep{FontRibera2014}.
From the fiducial $D^{fid}_{A}/s^{fid}$ and $H^{fid} \cdot s^{fid}$ values, we can easily calculate the corresponding errors $\sigma_{D_{A}/s}$ and $\sigma_{H \cdot s}$ from columns $2$ and $3$ in Table~6 of \citep{FontRibera2014}.

For our analysis, though, we will proceed just a bit differently and randomly pick up $D_{A}/s$ and $H \cdot s$ from a multivariate Gaussian centered on $D^{fid}_{A}/s^{fid}$ and $H^{fid} \cdot s^{fid}$, and with standard deviation proportional to $\sigma_{D_{A}/s}$ and $\sigma_{H \cdot s}$ and correlation
$r = 0.4$ \citep{SeoEisenstein2007}, in order to give to mock data an intrinsic dispersion closer to the real one. Of course, we cannot rely on the results from only one single random run; instead we realize $10^3$ random mock data sets, and we apply our algorithm to each of them. Our final result\textst{s} will be then a statistical output on an ensemble of possible models observationally compatible with our starting fiducial model. We point out here that this is the only step in our work where assuming a cosmological model is needed. This step is quite unavoidable in order to have a reference point to establish the goodness of our analysis, but it is a quite common procedure in forecast analysis; and it is a very general assumption. Moreover, our choice to test our method on a large number of data sets will greatly smear the effects of this initial input.

But in order to calculate the fiducial quantities in the context of VSL, we need an ansatz for $c(z)$. So far no definitive theoretical background exists for VSL. We have chosen to follow the approach summarized in \citep{Magueijo2003}, where a minimal coupling is assumed between matter and the field driving the change in the speed of light. This implies that there is no change in the continuity equation for each cosmological fluid, and in the Friedmann equation (i.e. in the observational $H(z)$), provided that the spatial curvature is zero. As we assume no spatial curvature, thus the only change occurs in the calculation of cosmological distances ($D_{A}$ and $s$), which involve the change in the integral:
\begin{equation}
\int^{z_{2}}_{z_{1}} \frac{c_{0}}{H(z')} \, d \, z' \rightarrow \int^{z_{2}}_{z_{1}} \frac{c(z')}{H(z')} \, d \, z' \; ,
\end{equation}
where the speed of light is no longer constant $(c_{0})$, but a function $c(z)$. At the present stage, there is no clear and reliable phenomenological expression for $c(z)$ \citep{VSL_observations}; we have chosen to work with the general theoretically-motivated expression from \citep{Magueijo2003}, $c(a) \propto c_{0} \left( 1+ a/a_{c} \right)^{n}$, where $a \equiv 1/(1+z)$ is the scale factor, and $a_{c}$ is the transition epoch from some $c(a) \neq c_{0}$ (at early times) to $c(a) \rightarrow c_{0}$ (at late times - now). Another possible ansatz is $c \propto c_{0} a^{n}$ \citep{VSL_theory}, but it is less flexible in order to (qualitatively) match both early and late times observations. We stress, however, that the choice of the $c(z)$ function is only needed to simulate some mock observational data with some intrinsic variation of $c$, and has no influence at all on the final results.

For our analysis, we have considered three scenarios: one with $c \equiv c_{0}$ constant; one with $a_{c} = 0.005$ and $n = -0.01$; and one with $a_{c} = 0.005$ and $n = -0.001$. The second case corresponds to a $\Delta c / c_{0} \approx 1 \%$ at $z \sim 1.5$; the third to a $\Delta c / c_{0} \approx 0.1 \%$.  In order to make the global dynamics of the Universe within these two VSL scenarios compatible with present data, we have to change the value of $\Omega_{m}$ (dimensionless matter density today). This is expected, because VSL can solve (also and/or partially) the dark energy problem: a higher speed of light in the past can mimic the effects of a dark energy component.
This effect results in a lower value for $\Omega_{DE}$ (dimensionless dark energy density today) or, equivalently, assuming no spatial curvature, in a larger value of $\Omega_{m}$. In the classical context of constant $c$, the chosen \textit{Planck} value is $\Omega_{m} = 0.308$. In order to arrange for the above assessed variations in $c$, in the second model of VSL, we need $\Omega_{m} = 0.380$, which corresponds to changes in $D_{A}$ which are $<0.5\sigma^{now}_{D_{A}}$ in the redshift range $[0,1]$ (for which we have data now), where $\sigma^{now}_{D_{A}}$ is the error from present surveys. In the third VSL model, we need $\Omega_{m} = 0.315$, corresponding to changes in $D_{A}$ of $<0.1\sigma^{now}_{D_{A}}$ in the same redshift range. Moreover, the changes in the sound horizon are $\lesssim \sigma_{s}$ in both cases, with $\sigma_{s}$ the error from the chosen \textit{Planck} fiducial model.

Once we have our mock data sets, $D_{A}/s$ and $H \cdot s$, and related errors, $\sigma_{D_{A}/s}$ and $\sigma_{H \cdot s}$, we reconstruct the underlying smooth functions using GPs. Following \citep{Seikel2012}, we apply a Markov Chain at each mock data set in order to find the GPs parameters that optimize the reconstruction of $D_{A}/s$ and $H \cdot s$. Then, we evaluate the GPs output function on a $\Delta z = 0.01$ redshift grid. This grid (ten times finer than the \textit{Euclid} data given in Table~6 of \citep{FontRibera2014}) is useful to implement a numerical algorithm to calculate $z_{M}$ for each simulation. We fit $D_{A}/s$ and $c_{0}/(H \cdot s)$ with a high order polynomial in the redshift range $[1.,2.]$ and then we find $z_{M}$ analytically with its related error $\sigma_{z_{M}}$.

Finally, we calculate the quantity $D_{A}(z) \cdot H(z)/c_{0}$ (note that the exact value of the sound horizon is non influential at this step) from the GPs reconstructed data sets and, using $z_{M}\pm\sigma_{z_{M}}$ we can constrain the speed of light. Given our choice to normalize $D_{A}(z) \cdot H(z)$ with $c_{0}$, in the context of constant speed of light we expect to find $D_{A}(z_{M}) \cdot H(z_{M})/c_{0} \approx 1$ with some error, recalling that in VSL theories it can be different from $1$.

\begin{figure}[htbp]
\includegraphics[width=8.3cm]{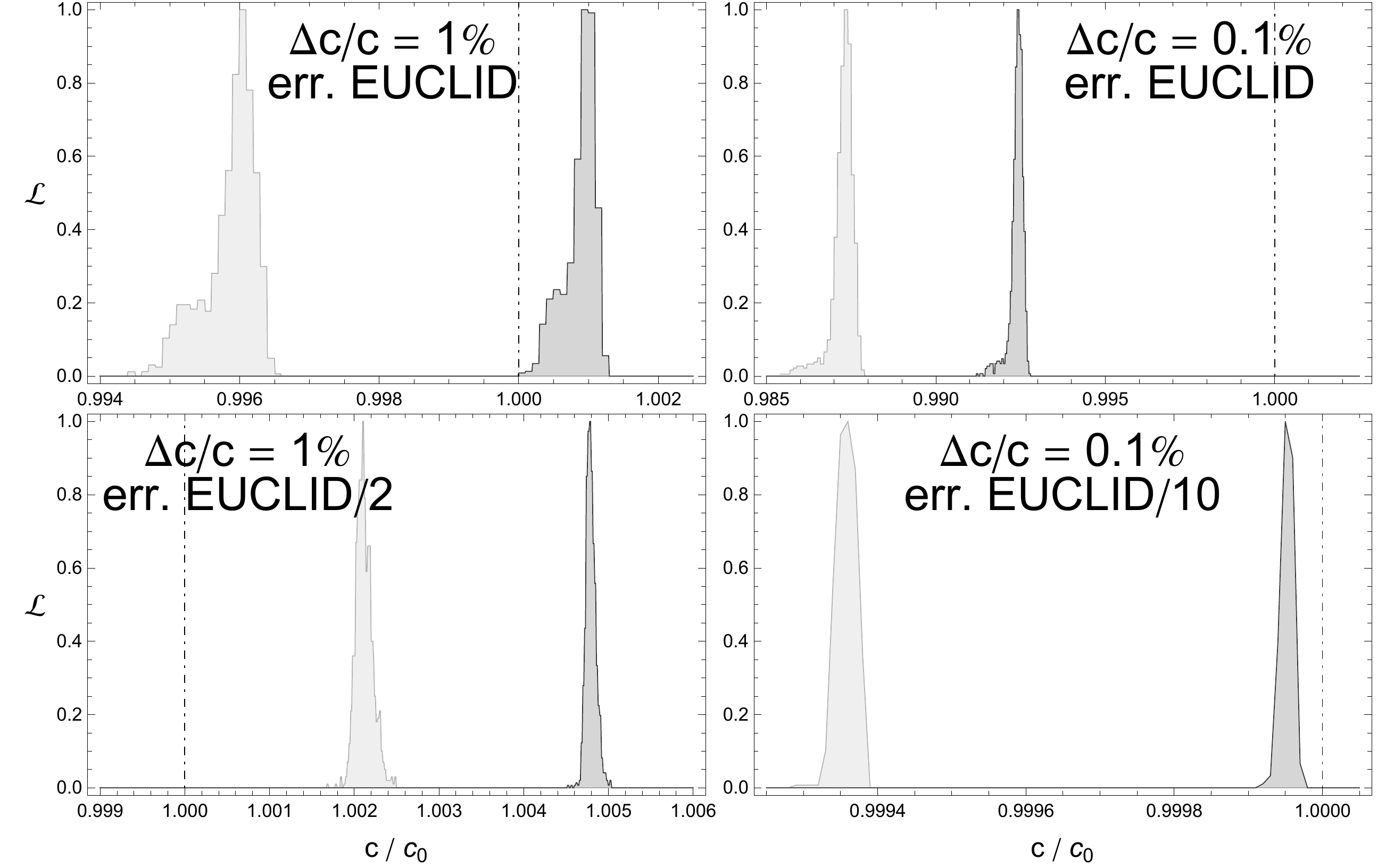}
\caption{Probability distribution of $c(z_{M})/c_{0} - 1\sigma_{c(z_{M})/c_{0}}$ (dark grey) and $c(z_{M})/c_{0} - 2\sigma_{c(z_{M})/c_{0}}$ (light grey) from $10^3$ simulations in different survey configurations. Vertical black dot-dashed line is for $c(z_{M}) = c_{0}$.}\label{fig:histograms}
\end{figure}


\textit{Results and discussion.} As a preliminary step, we have tested our algorithm with constant $c = c_{0}$. We obtain: $z_{M} = 1.592^{+0.043}_{-0.039}$ (the fiducial input value is $z_{M} =1.596$) and $c = 1 \pm 0.009$ \citep{footnote2}. We assess that our method works quite well, as we are able to recover the input model with a very good accuracy.

The main question now is: \textit{is this accuracy enough to detect possible VSL?} To answer, we consider the VSL model with a $1\%$ variation
in $c$. Results are: $z_{M} = 1.528^{+0.038}_{-0.036}$ (fiducial input is $z_{M} =1.532$) and $c(z_{M}) = 1.00925 \pm 0.00831$ (fiducial input: $c(z_{M}) = 1.00926$). If we consider the lower $1\sigma$ limit for $c(z_{M})$, we have the following results (from our $10^3$ simulations):
+$\langle c(z_{M})-1\sigma_{c(z_{M})}\rangle = 1.00094^{+0.00014}_{-0.00033}$. Thus, \textit{Euclid} will be able to detect (if any) a VSL of $\sim 1\%$ at $1\sigma$ level (top left panel in Fig.~\ref{fig:histograms}).

When we apply our method to the model corresponding to a $0.1\%$ variation in $c$ we have: $z_{M} = 1.584^{+0.042}_{-0.039}$ (fiducial input is $z_{M} =1.589$) and $c(z_{M}) = 1.00095 \pm 0.00852$ (fiducial input: $c(z_{M}) = 1.00094$). The lower $1\sigma$ limit for $c(z_{M})$ is $\langle c(z_{M})-1\sigma_{c(z_{M})}\rangle = 0.99243^{+0.00013}_{-0.00016}$. It is clear that in this case \textit{Euclid} will fail in detecting VSL at even $1\sigma$ level (top right panel in Fig.~\ref{fig:histograms}).

Having assumed that already-planned surveys will not be able to detect a VSL smaller than $1\%$, we have explored whether there is any chance for BAO to perform better. Building a reliable BAO survey in all possible details has many constructive difficulties and it is out of the purpose of this letter.
We have thus carried out a naive ``rule-of-thumb'' analysis: we have assumed a \textit{Euclid}-style survey
(i.e., with the same redshift range and bins as \textit{Euclid}), but with a better performance, quantified as smaller errors on $D_{A}$ and $H$.
We have found that reducing the errors by half, for example, will make it possible to detect $1\%$ VSL at even $2\sigma$ (bottom left panel in Fig.~\ref{fig:histograms}): $z_{M} = 1.531^{+0.021}_{-0.020}$, $c(z_{M}) = 1.00926 \pm 0.00447$, and $\langle c(z_{M})-2\sigma_{c(z_{M})}\rangle = 1.00211^{+0.00011}_{-0.00008}$. On the other hand, no significant improvement would be obtained for the $0.1\%$ case. In order to start to spot something interesting in this case, we would need errors on the BAO mode at least $10$ times smaller than the expected \textit{Euclid} ones. With such low errors, we would have: $z_{M} = 1.5888^{+0.0048}_{-0.0047}$, $c(z_{M}) = 1.00095 \pm 0.00099$, and $\langle c(z_{M})-1\sigma_{c(z_{M})}\rangle = 0.999952\pm 0.000009$. Thus, such a survey would be at the border of the detectable limit of a $0.1\%$ VSL (bottom right panel in Fig.~\ref{fig:histograms}).

Limiting the VSL detection to $1\%$ or, in a quite optimistic case, to $0.1\%$, might be problematic: while there is no cosmological measurement of $c$
which can be used as a comparison tool, we have many measurements of another quantity strictly related to $c$, the fine-structure constant $\alpha \equiv e^{2}/ (\hbar \, c)$ ($e$ the electron charge). There are many observations which are compatible with varying $\alpha$ \citep{Uzan2011,alpha}. But these variations (if real) are always very small, at least $<10^{-4}$. From its definition it is easy to check that, if the other parameters involved in its definition are assumed to be constant, then $\Delta \alpha / \alpha = - \Delta c / c_{0}$. Thus, we would expect a variation for $c$ of the same-order. In principle, a large variation in $c$ might be compatible with such smaller variation in $\alpha$ if also the other parameters are allowed to vary. But in this case we would have an unpleasant ``fine-tuning'' problem, because in order to accommodate such small variation in $\alpha$, we
would need a larger variation from each of the other parameters to compensate each other. Thus, assuming such measurements of variation of $\alpha$ are correct, it is natural to expect the same order of variation for $c$. 
If we consider other cosmological probes, like the Cosmic Microwave Background (CMB), we have a detection of $\Delta \alpha / \alpha \sim 0.4 \%$ \citep{Planck_26}.
But this constraint is plagued by a strong degeneracy with the cosmological parameter $H_0$ (current expansion rate), and is obtained joining CMB with BAO and adding a prior on $H_{0}$.
From this point of view, we would like to stress that \textit{even our ``pessimistic-scenario'' of VSL detection from BAO is competitive with CMB, and it
would be obtained without any assumption on other possible cosmological parameters}.

Given such results and arguments, one last question arises: \textit{is it technically possible to achieve such small errors for BAO measurements and thus be able to measure finer variations of $c$?} From a quantitative point of view, the answer is not easy and would involve many technical problems. Qualitatively, we feel confident that this limit is within the reach of future observations. In \citep{EUCLID_1} (Fig.~2.21) and in \citep{FontRibera2014} (Fig.~3), the observational errors from many on-going and planned future surveys are shown.
It can be easily seen that \textit{Euclid} (expected) errors will be one tenth of the errors from an already completed
survey like the WiggleZ Dark Energy Survey \citep{WiggleZ_1,WiggleZ_2}, so this level of improvement is possible. Moreover,
it is also clear that ground based telescopes like the Dark Energy Spectroscopic Instrument (DESI) \citep{DESI} and the Square Kilometer Array (SKA) \citep{SKA} are almost as competitive as space ones like \textit{Euclid} and the Wide-Field Infrared Survey Telescope (\textit{WFIRST}) \citep{WFIRST}, so that in the future it will surely be possible to further improve space-based surveys and obtain better constraints.

There are also many other ways in which such constraints might be improved: \textit{1.} BAO have much smaller intrinsic dispersion with respect to other probes, like Type Ia Supernovae (see Fig.~12 in \citep{WiggleZ_1}); our $10^{3}$ simulations fall in a range larger than the $1\sigma$ interval expected from \textit{Euclid} observations (see Fig.~\ref{fig:histograms}), so that our results can be considered as a upper-pessimistic-level forecast; \textit{2.} we have chosen one particular reconstruction method, GPs, which is very powerful but, possibly, better methods might be employed when working with raw BAO data at an even more preliminary steps than the final-user one; 
\textit{3.} as we have described above, $z_{M}$ lies in the redshift range $[1.4,1.8]$ for many different cosmological models, and \textit{Euclid} is not optimized in this range; its best errors lie around $z \approx 1$; a survey like \textit{WFIRST} would be better designed for our scope, because the largest sensitivity is exactly around $\approx 1.5-1.6$, where the maximum redshift is very likely located.

\begin{figure}[htbp]
\includegraphics[width=8.5cm]{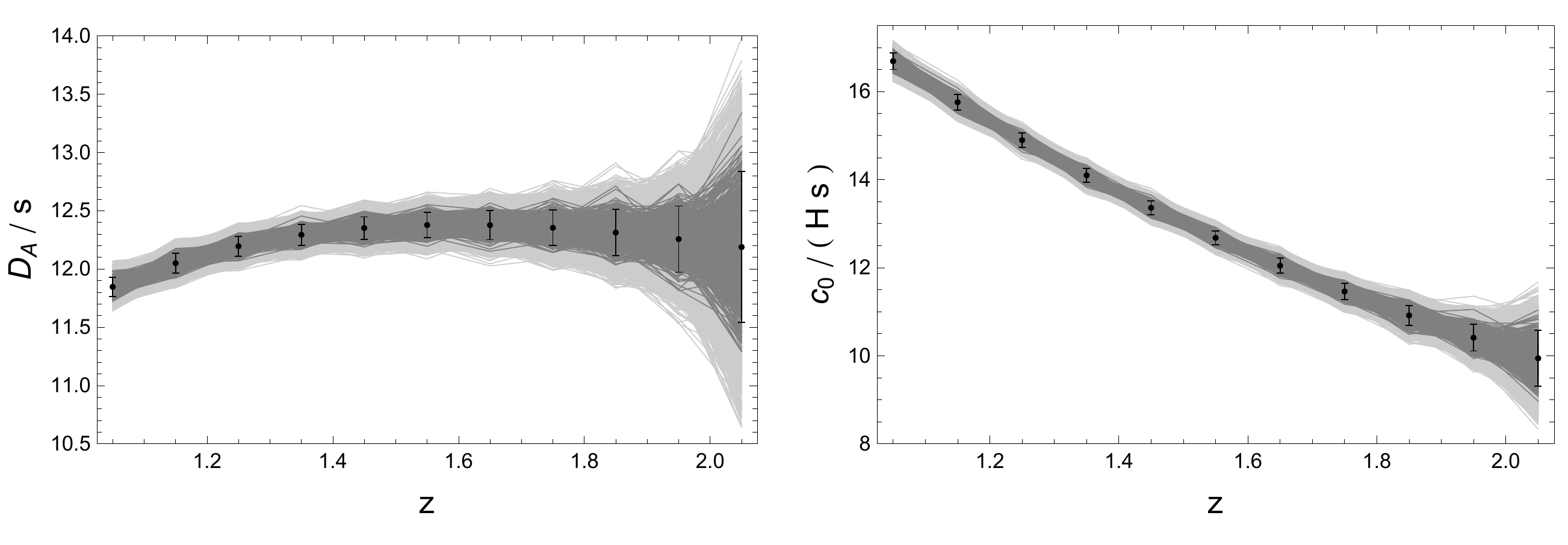}
\caption{$10^3$ simulations: black points - fiducial model; dark grey - simulations; light grey - $1\sigma$ from simulations.}\label{fig:simulations}
\end{figure}

\section{Acknowledgements}

The research of M.P.D. and V.S. was financed by the National Science Center Grant DEC-2012/06/A/ST2/00395. R.L. and V.S. were supported by the Spanish Ministry of Economy and Competitiveness through research projects FIS2010-15492 and Consolider EPI CSD2010-00064. R.L.  was also  financially aided by the University of the Basque Country UPV/EHU under program UFI 11/55 and by the Basque Government through research project GIC12/66.

\vfill

\begin{thebibliography}{99}
\bibitem{footnote1}
We are aware that the $w+w_{a}$ is only one of the many dark energy models available, but it is somewhat used as a ``reference'' model in the literature. Moreover, large errors on its parameters, in particular on $w_{a}$, make us confident to have explored a very large set of cosmological scenarios, so that our estimation for the range of $z_{M}$ is highly conservative.
\bibitem{footnote2}
Note that, given our choice to work with an ensemble of $10^3$ possible observational sets, all the numbers given have to be read as: best value $\equiv$ mean of the best value from all $10^3$ simulations; error $\equiv$ mean of the error from all $10^{3}$ simulations.
\end{thebibliography}
\end{document}